\documentclass[aps,prl,superscriptaddress,showpacs,floatfix,twocolumn]{revtex4}

\usepackage{graphicx}	% Include figure files

\begin{document}

\title{Production of $\omega$ mesons at Large Transverse Momenta \\ 
in $p+p$ and $d$+Au Collisions at $\sqrt{s_{NN}}=200$\,GeV}

\newcommand{\abilene}{Abilene Christian University, Abilene, TX 79699, USA}
\newcommand{\acadsin}{Institute of Physics, Academia Sinica, Taipei 11529, Taiwan}
\newcommand{\banaras}{Department of Physics, Banaras Hindu University, Varanasi 221005, India}
\newcommand{\barc}{Bhabha Atomic Research Centre, Bombay 400 085, India}
\newcommand{\bnl}{Brookhaven National Laboratory, Upton, NY 11973-5000, USA}
\newcommand{\caucr}{University of California - Riverside, Riverside, CA 92521, USA}
\newcommand{\ciae}{China Institute of Atomic Energy (CIAE), Beijing, People's Republic of China}
\newcommand{\cns}{Center for Nuclear Study, Graduate School of Science, University of Tokyo, 7-3-1 Hongo, Bunkyo, Tokyo 113-0033, Japan}
\newcommand{\colorado}{University of Colorado, Boulder, CO 80309, USA}
\newcommand{\columbia}{Columbia University, New York, NY 10027 and Nevis Laboratories, Irvington, NY 10533, USA}
\newcommand{\dapnia}{Dapnia, CEA Saclay, F-91191, Gif-sur-Yvette, France}
\newcommand{\debrecen}{Debrecen University, H-4010 Debrecen, Egyetem t{\'e}r 1, Hungary}
\newcommand{\elte}{ELTE, E{\"o}tv{\"o}s Lor{\'a}nd University, H - 1117 Budapest, P{\'a}zm{\'a}ny P. s. 1/A, Hungary}
\newcommand{\fsu}{Florida State University, Tallahassee, FL 32306, USA}
\newcommand{\gsu}{Georgia State University, Atlanta, GA 30303, USA}
\newcommand{\hiroshima}{Hiroshima University, Kagamiyama, Higashi-Hiroshima 739-8526, Japan}
\newcommand{\ihepprot}{IHEP Protvino, State Research Center of Russian Federation, Institute for High Energy Physics, Protvino, 142281, Russia}
\newcommand{\illuiuc}{University of Illinois at Urbana-Champaign, Urbana, IL 61801, USA}
\newcommand{\isu}{Iowa State University, Ames, IA 50011, USA}
\newcommand{\jinrdubna}{Joint Institute for Nuclear Research, 141980 Dubna, Moscow Region, Russia}
\newcommand{\kek}{KEK, High Energy Accelerator Research Organization, Tsukuba, Ibaraki 305-0801, Japan}
\newcommand{\kfki}{KFKI Research Institute for Particle and Nuclear Physics of the Hungarian Academy of Sciences (MTA KFKI RMKI), H-1525 Budapest 114, POBox 49, Budapest, Hungary}
\newcommand{\korea}{Korea University, Seoul, 136-701, Korea}
\newcommand{\kurchatov}{Russian Research Center ``Kurchatov Institute", Moscow, Russia}
\newcommand{\kyoto}{Kyoto University, Kyoto 606-8502, Japan}
\newcommand{\labllr}{Laboratoire Leprince-Ringuet, Ecole Polytechnique, CNRS-IN2P3, Route de Saclay, F-91128, Palaiseau, France}
\newcommand{\lawllnl}{Lawrence Livermore National Laboratory, Livermore, CA 94550, USA}
\newcommand{\losalamos}{Los Alamos National Laboratory, Los Alamos, NM 87545, USA}
\newcommand{\lpc}{LPC, Universit{\'e} Blaise Pascal, CNRS-IN2P3, Clermont-Fd, 63177 Aubiere Cedex, France}
\newcommand{\lund}{Department of Physics, Lund University, Box 118, SE-221 00 Lund, Sweden}
\newcommand{\muenster}{Institut f\"ur Kernphysik, University of Muenster, D-48149 Muenster, Germany}
\newcommand{\myongji}{Myongji University, Yongin, Kyonggido 449-728, Korea}
\newcommand{\nagasaki}{Nagasaki Institute of Applied Science, Nagasaki-shi, Nagasaki 851-0193, Japan}
\newcommand{\newmex}{University of New Mexico, Albuquerque, NM 87131, USA }
\newcommand{\nmsu}{New Mexico State University, Las Cruces, NM 88003, USA}
\newcommand{\ornl}{Oak Ridge National Laboratory, Oak Ridge, TN 37831, USA}
\newcommand{\orsay}{IPN-Orsay, Universite Paris Sud, CNRS-IN2P3, BP1, F-91406, Orsay, France}
\newcommand{\peking}{Peking University, Beijing, People's Republic of China}
\newcommand{\pnpi}{PNPI, Petersburg Nuclear Physics Institute, Gatchina, Leningrad region, 188300, Russia}
\newcommand{\riken}{RIKEN (The Institute of Physical and Chemical Research), Wako, Saitama 351-0198, JAPAN}
\newcommand{\rikjrbrc}{RIKEN BNL Research Center, Brookhaven National Laboratory, Upton, NY 11973-5000, USA}
\newcommand{\saopaulo}{Universidade de S{\~a}o Paulo, Instituto de F\'{\i}sica, Caixa Postal 66318, S{\~a}o Paulo CEP05315-970, Brazil}
\newcommand{\seoulnat}{System Electronics Laboratory, Seoul National University, Seoul, South Korea}
\newcommand{\stonybrkc}{Chemistry Department, Stony Brook University, Stony Brook, SUNY, NY 11794-3400, USA}
\newcommand{\stonycrkp}{Department of Physics and Astronomy, Stony Brook University, SUNY, Stony Brook, NY 11794, USA}
\newcommand{\subatech}{SUBATECH (Ecole des Mines de Nantes, CNRS-IN2P3, Universit{\'e} de Nantes) BP 20722 - 44307, Nantes, France}
\newcommand{\tenn}{University of Tennessee, Knoxville, TN 37996, USA}
\newcommand{\titech}{Department of Physics, Tokyo Institute of Technology, Oh-okayama, Meguro, Tokyo 152-8551, Japan}
\newcommand{\tsukuba}{Institute of Physics, University of Tsukuba, Tsukuba, Ibaraki 305, Japan}
\newcommand{\vandy}{Vanderbilt University, Nashville, TN 37235, USA}
\newcommand{\waseda}{Waseda University, Advanced Research Institute for Science and Engineering, 17 Kikui-cho, Shinjuku-ku, Tokyo 162-0044, Japan}
\newcommand{\weizmann}{Weizmann Institute, Rehovot 76100, Israel}
\newcommand{\yonsei}{Yonsei University, IPAP, Seoul 120-749, Korea}
\affiliation{\abilene}
\affiliation{\acadsin}
\affiliation{\banaras}
\affiliation{\barc}
\affiliation{\bnl}
\affiliation{\caucr}
\affiliation{\ciae}
\affiliation{\cns}
\affiliation{\colorado}
\affiliation{\columbia}
\affiliation{\dapnia}
\affiliation{\debrecen}
\affiliation{\elte}
\affiliation{\fsu}
\affiliation{\gsu}
\affiliation{\hiroshima}
\affiliation{\ihepprot}
\affiliation{\illuiuc}
\affiliation{\isu}
\affiliation{\jinrdubna}
\affiliation{\kek}
\affiliation{\kfki}
\affiliation{\korea}
\affiliation{\kurchatov}
\affiliation{\kyoto}
\affiliation{\labllr}
\affiliation{\lawllnl}
\affiliation{\losalamos}
\affiliation{\lpc}
\affiliation{\lund}
\affiliation{\muenster}
\affiliation{\myongji}
\affiliation{\nagasaki}
\affiliation{\newmex}
\affiliation{\nmsu}
\affiliation{\ornl}
\affiliation{\orsay}
\affiliation{\peking}
\affiliation{\pnpi}
\affiliation{\riken}
\affiliation{\rikjrbrc}
\affiliation{\saopaulo}
\affiliation{\seoulnat}
\affiliation{\stonybrkc}
\affiliation{\stonycrkp}
\affiliation{\subatech}
\affiliation{\tenn}
\affiliation{\titech}
\affiliation{\tsukuba}
\affiliation{\vandy}
\affiliation{\waseda}
\affiliation{\weizmann}
\affiliation{\yonsei}
\author{S.S.~Adler}	\affiliation{\bnl}
\author{S.~Afanasiev}	\affiliation{\jinrdubna}
\author{C.~Aidala}	\affiliation{\columbia}
\author{N.N.~Ajitanand}	\affiliation{\stonybrkc}
\author{Y.~Akiba}	\affiliation{\kek} \affiliation{\riken}
\author{A.~Al-Jamel}	\affiliation{\nmsu}
\author{J.~Alexander}	\affiliation{\stonybrkc}
\author{K.~Aoki}	\affiliation{\kyoto}
\author{L.~Aphecetche}	\affiliation{\subatech}
\author{R.~Armendariz}	\affiliation{\nmsu}
% ...
\collaboration{PHENIX Collaboration} \noaffiliation

\date{\today}

\begin{abstract}

The PHENIX experiment at RHIC has measured the invariant cross section for 
$\omega$-meson production at mid-rapidity in the transverse momentum range 
$2.5 < p_{\rm T} < 9.25$\,GeV/$c$ in $p+p$ and $d$+Au collisions at 
$\sqrt{s_{NN}}=200$\,GeV.  Measurements in two decay channels 
($\omega\rightarrow\pi^0\pi^+\pi^-$ and $\omega\rightarrow\pi^0\gamma$) 
yield consistent results, and the reconstructed $\omega$ mass agrees with 
the accepted value within the $p_{\rm T}$ range of the measurements.  The 
$\omega/\pi^0$ ratio is found to be $0.85 \pm 0.05^{\rm stat} \pm 
0.09^{\rm sys}$ in $p+p$ and $0.94 \pm 0.08^{\rm stat} \pm 0.12^{\rm sys}$ 
in $d$+Au collisions, independent of $p_{\rm T}$.  The nuclear 
modification factor $R^{\omega}_{d{\rm A}}$ is $1.03 \pm 0.12^{\rm stat} 
\pm 0.21^{\rm sys}$ and $0.83 \pm 0.21^{\rm stat} \pm 0.17^{\rm sys}$ in 
minimum bias and central (0-20\%) $d$+Au collisions, respectively.

\end{abstract}

	
\maketitle

%\section{INTRODUCTION}

Cross sections at large transverse momentum ($p_{\rm T}$) for products of
``hard'' point-like processes ({\em e.g.}, inclusive hadrons, jets,
direct $\gamma$'s, and heavy flavor) in high energy hadron collisions
are well described in perturbative Quantum Chromodynamics
(pQCD)~\cite{geist90}.  As a result they are considered to be 
well-calibrated
probes of small-distance QCD phenomena.  When going from $p+p$ to $p$(or 
$d$)+A
and A+A collisions, deviations from cross section scaling with respect
to the number of binary N+N collisions provide information
on cold nuclear matter effects such as initial state energy
loss~\cite{loss}, shadowing~\cite{shadow} and hot nuclear matter effects
such as in-medium energy loss~\cite{aaa}, increasing importance of
production via recombination~\cite{recomb}, and modifications to the QCD
vacuum~\cite{vacuum}.

Production of $\omega$ mesons at high-$p_{\rm T}$ is especially 
interesting.  The $\omega$ and $\pi^0$ are vector and pseudoscalar mesons, 
respectively.  The $\omega/\pi^0$ ratio carries information about 
probabilities of corresponding spin states to be produced in 
hadronization.  Furthermore, $\omega$ mesons can be used as a probe of the 
nuclear medium since a significant fraction of them produced in A+A or 
$p$($d$)+A collisions will decay inside the produced medium ($c\tau = 
23.8$\,fm) possibly leading to changes of the mass and/or width of the 
$\omega$ with respect to their values in vacuum.

%\section{SETUP}

% We report measurements of $\omega$ meson production in 
% $d$+Au and $p+p$ collisions at $\sqrt{s_{NN}}=200$\,GeV.  
The PHENIX~\cite{phenix} experiment at RHIC has a unique capability to 
measure 
both neutral and charged products of A+A and $p+p$ collisions at 
very high event rates.  The two central arm spectrometers each cover 
90$^{\circ}$ in azimuth and $\pm0.35$ in pseudorapidity.  The 
Electromagnetic Calorimeter (EMCal), with energy resolution $\sigma/E = 
8.1$\%$/\sqrt{E{\rm (GeV)}}\oplus 2.1$\%, is used to reconstruct 
$\gamma$'s and $\pi^0$'s.  For charged particle reconstruction two layers 
of Pad Chambers provide 3D pattern recognition and fake track rejection, 
and a Drift Chamber gives momentum resolution $\sigma/p_{\rm T} = 
0.7$\%$\sqrt{p_{\rm T}(GeV/c)} \oplus 1.1$\%.  Beam-Beam counters (BBC) were 
used to provide minimum bias trigger and to determine the collision 
vertex.  The minimum bias trigger cross sections measured by the BBC are 
$23.0 \pm 2.2$\,mb in $p+p$~\cite{ppg24} and $1.99 \pm 0.10$\,b in 
$d$+Au~\cite{dau_totalxsec} collisions.  In $d$+Au collisions the BBC were 
also used to separate events into centrality classes as explained 
in~\cite{ppg55}.  High $p_{\rm T}$ online triggers are implemented by adding 
together amplitudes in $4\times4$ adjacent EMCal towers and comparing them 
to a threshold of 1.4\,GeV in $p+p$ and 2.4\,GeV in $d$+Au.  The trigger 
could be fired by one or more photons coming from $\omega$ decay final 
states including $3\gamma$ or $2\pi 2\gamma$.

The data was collected by the PHENIX experiment during RHIC Run3. 
After selecting good runs and cutting on the collision vertex
($|z| < 30$\,cm) we analyzed approximately $4.6\times10^7$ and
$2.1\times10^7$ high $p_{\rm T}$ trigger events, corresponding to a total integrated
luminosity of 0.22\,pb$^{-1}$ and 1.5\,nb$^{-1}$ in the $p+p$ and
$d$+Au collision systems, respectively.

%\section{ANALYSIS}
Although $\omega$-mesons are relatively abundant in high energy
hadronic collisions ($\omega/\pi^0 \approx 1$ at high $p_{\rm T}$), their
measurement is challenging due to the multi-particle final states of
the main decay channels and the combinatorial background associated
with their reconstruction. 

\begin{figure}[tbh]
\includegraphics[width=1.0\linewidth]{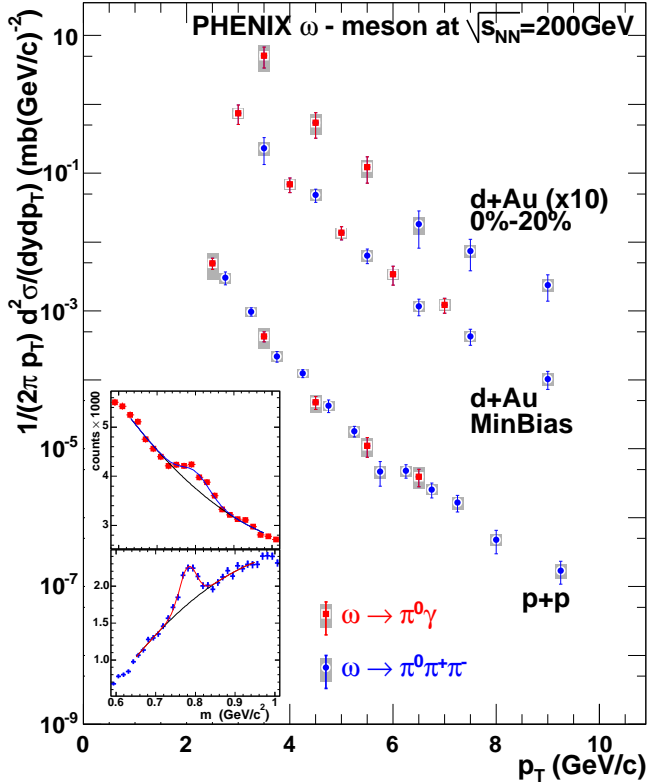}
\caption{Invariant cross section of $\omega$ production in $p+p$ and
$d$+Au collisions at $\sqrt{s_{NN}}=200$\,GeV measured in $\omega
\rightarrow \pi^{0}\pi^{+}\pi^{-}$ and $\omega \rightarrow
\pi^{0}\gamma$ decay channels.  Bars and boxes represent statistical
and systematic errors, respectively.  Fits to invariant mass
distributions of $\pi^{0}\gamma$ (top) and $\pi^{0}\pi^{+}\pi^{-}$ (bottom) are shown
in the insert.
\label{fig:spectra}}
\end{figure}

The procedure used to measure $\omega\rightarrow\pi^{0}\pi^{+}\pi^{-}$
is the same as used to measure $\eta\rightarrow\pi^{0}\pi^{+}\pi^{-}$~\cite{ppg55}. 
In case of $\omega$ one shall account for a wider peak and
different phase space density of the 3-body
decay~\cite{dalitz1,dalitz2}. 
The analysis procedure for the photonic
decay mode, $\omega\rightarrow\pi^{0}\gamma\rightarrow 3\gamma$, is
very similar to earlier PHENIX measurements of other mesons with
photonic decay modes, $\pi^0,\eta\rightarrow
2\gamma$~\cite{ppg24,ppg14,ppg51}.  For both $\omega$ decay modes
studied the first step is to reconstruct $\pi^0$ candidates by
combining photon pairs and applying a $p_{\rm T}$ dependent cut around
the mass of the $\pi^0$.  The R.M.S. of the $\pi^{0}$ peak varies 
with $p_{\rm T}$ from 8\,MeV to 13\,MeV.  Candidates (which include 
combinatorial
background) are combined with a third photon for
$\omega\rightarrow\pi^0\gamma$ or with two unidentified charged tracks
(assumed to be $\pi$-mesons) for $\omega\rightarrow\pi^0\pi^+\pi^-$. Raw yields are extracted by
fitting the $p_{\rm T}$ slices of the invariant mass
distribution as in the insert panel in Fig.~\ref{fig:spectra}. 
The signal to background ratio (S/B) of $\pi^0\pi^+\pi^-$ decay channel 
is 1:7 at low $p_{\rm T}$ and grows to 1:1 with $p_{\rm T}$ in $p+p$
collisions.  In $d$+Au collisions it starts at 1:20 and grows with
$p_{\rm T}$ to 1:2.  For the $\pi^{0}\gamma$ channel S/B is 1.5-2 times worse.
Corrections for acceptance, trigger efficiency, and
analysis cuts are described in detail in~\cite{ppg55}.  Further details
about the analysis procedures can be found in~\cite{victor}.

We classify systematic error sources as type A (point-to-point
uncorrelated, which can move each point independently), type B
(point-to-point correlated, which can move points coherently, but
potentially by different relative amounts), and type C (global, which
move all points by the same relative amount).  These errors are
summarized in Table~\ref{tab:syserr} for the different decay modes and
collision systems.  Major contributors include signal extraction (type
A), online high-$p_{\rm T}$ trigger efficiency (type B) and the total
cross-section measurement (type C).  The uncertainty on the signal
extraction, which is the dominant source of systematic error, is
relatively large due to the fact that correlations in the triggered
sample rendered background subtraction via event mixing impossible, and
therefore the unknown shape of the underlying background had to be
accounted for in the peak fit.  This error is estimated based on
variation of analysis cuts and fitting procedures~\cite{ppg55}.

%%%%%%%%%%%%%%%%%%%%%%%%%%%%%%%%%%%%%%%%%%%%%%%% Table I
\begin{table}[ht]
\caption{\label{tab:syserr} Systematic errors for the $\omega$
  production cross section for the two decay channels and the two
  collision systems analyzed.  Error types are described in the
  text.  Values with a range indicate variation of the systematic error
  over the $p_{\rm T}$ range of the measurement.}
\begin{ruledtabular}\begin{tabular}{ccccc}
& \multicolumn{2}{c}{$\omega\rightarrow\pi^0\pi^+\pi^-$} 
& \multicolumn{2}{c}{$\omega\rightarrow\pi^0\gamma$} \\
Type  & $p+p$ & $d$+Au & $p+p$   & $d$+Au            \\ 
\hline 
A     & 7-20\%  & 10-15\%  &   25-40\% &     10\%          \\
B     & 8-10\%  & 11-14\%  & 5.8-9.2\% & 7.4-11\%          \\
C     &   11\%  &   9.4\%  &      13\% &     11\%          \\
\end{tabular}\end{ruledtabular}
\end{table}

%\section{RESULTS}

Figure~\ref{fig:spectra} shows 
the invariant cross sections for $\omega$ production in $p+p$, minimum
bias $d$+Au, and central (0-20\%) $d$+Au collisions at $\sqrt{s_{NN}}
= 200$\,GeV, as a function of $p_{\rm T}$.
The $p_{\rm T}$ range ($2.5 < p_{\rm T} < 9.25$\,GeV/$c$) 
is limited by statistics at high $p_{\rm T}$, and by
decreasing detector acceptance and trigger efficiency at low
$p_{\rm T}$.  The results for the two decay modes, which involve different 
kinematics,
detector acceptance, and efficiency corrections, agree very well.

%%%%%%%%%%%%%%%%%%%%%%%%%%%%%%%%%%%%%%%%%%%%%%%%%%%%%%%%%%%%%%  Fig. 1

\begin{figure} [tbh]
\includegraphics[width=1.0\linewidth]{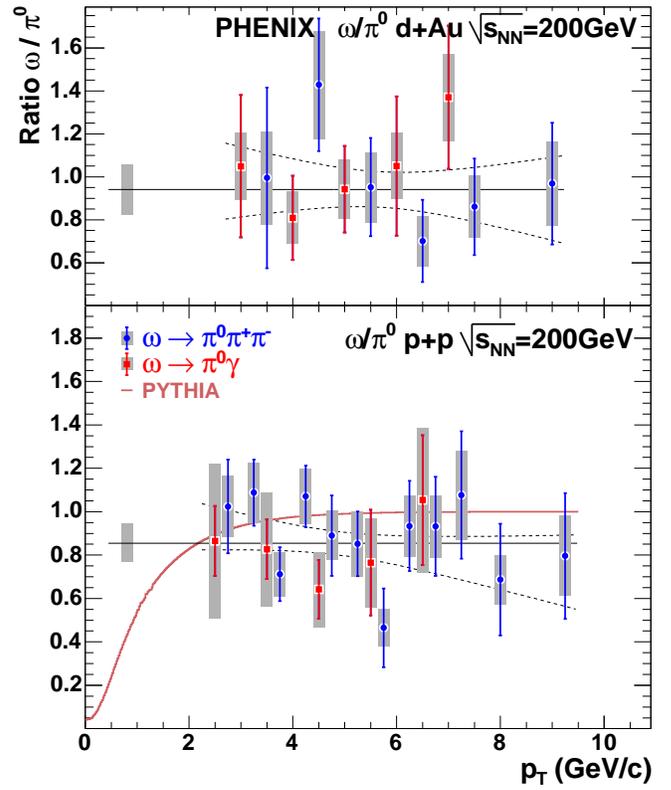}
\caption{Measured $\omega/\pi^{0}$ vs $p_{\rm T}$ in 
(upper panel) $d$+Au and 
(lower panel) $p+p$ collisions at $\sqrt{s_{NN}}=200$\,GeV.   
Straight lines
show fits to a constant for each collision system.  The boxes at
the left edge of the constant fit lines show the systematic error on the
data averaged over $p_{\rm T}$.  Dashed lines show values within $1\sigma$ of
the best linear fits to the data.  The PYTHIA prediction~\cite{pythia} for 
$p+p$ collisions at
$\sqrt{s}=200$\,GeV is shown with a solid curve in the bottom panel.\label{fig:ratio}}
\end{figure}

Figure~\ref{fig:ratio} shows the results for the ratio of vector
to pseudoscalar meson production ($\omega/\pi^{0}$) in $d$+Au
and $p+p$ collisions at $\sqrt{s_{NN}}=200$\,GeV.  For the
denominator we use inclusive $\pi^{0}$ yields measured by
PHENIX~\cite{ppg24,ppg28}.  Ratios in both systems are consistent with
unity over all measured $p_{\rm T}$.  Fits to a constant yield $\omega/\pi^{0} 
= 0.85
\pm 0.05^{\rm stat}  \pm 0.09^{\rm sys}$ and $0.94 \pm 0.08
   ^{\rm stat} \pm 0.12^{\rm sys}$ in
$p+p$ and $d$+Au collisions, respectively.  Fits assuming linear $p_{\rm T}$
dependence have slopes consistent with zero.  At high $p_{\rm T}$ the
PYTHIA~\cite{pythia} prediction for this ratio in $p+p$ collisions at
$\sqrt{s}=200$\,GeV is consistent with, but slightly higher
than, the measurement.

%%%%%%%%%%%%%%%%%%%%%%%%%%%%%%%%%%%%%%%%%%%%%%%%%%%%%%%%%%%%%%  Fig. 2

The R-806 experiment at ISR  measured $\omega/\pi^{0}$ in $p+p$ collisions at
$\sqrt{s}=62$\,GeV~\cite{isr} and found this ratio to be
$0.87\pm0.17$ over $3.5 < p_{\rm T} < 7$\,GeV/$c$.  The E706 experiment
measured $\omega/\pi^{0}$ in $\pi^{-}$+Be collisions at $\sqrt{s_{\pi
N}}=31$\,GeV~\cite{e706} and found values consistent with the results
presented in this paper.  The $\omega/\pi^{0}$ ratios measured in hadronic 
interactions by three
experiments at three different energies between 31 and 200\,GeV, are
the same within the errors.

Several LEP experiments~\cite{L3,ALEPH,OPAL} have measured $\omega$
production in $e^++e^-$ collisions at $\sqrt{s}=91.2$\,GeV and
presented results as a function of $x_{p} =
2p_{\omega}/\sqrt{s}$.  This is not a well-defined quantity in hadronic
interactions.  However, one can compare the $\omega/\pi^{0}$ ratio at
large values of $p_{\rm T}$ and $x_p$.  Following the procedure for
handling LEP data described in~\cite{ppg55} one finds that for $x_p >
0.5$, the largest value for which statistically significant LEP data
is available, the ratio has grown to approximately 0.7, close to the
measurements in hadronic interactions.

\begin{figure}[tbh]
\includegraphics[width=1.0\linewidth]{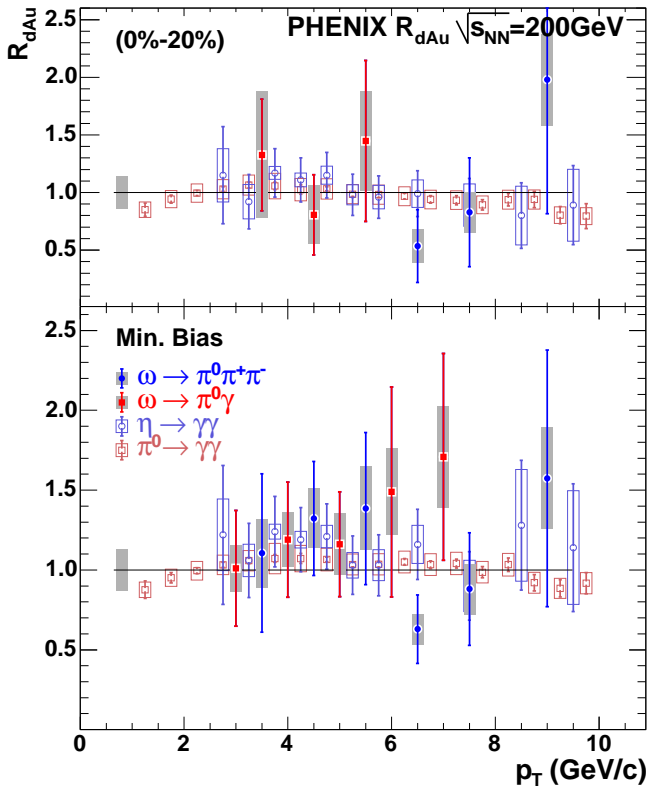}
\caption{\label{fig:rda}
Measured $R_{d{\rm A}}$ vs $p_{\rm T}$ for neutral mesons in $d$+Au 
collisions at $\sqrt{s_{NN}}=200$\,GeV for (upper panel) 0\%-20\% central 
and (lower panel) minimum bias.  Values for $\pi^{0}$'s and $\eta$'s are 
from~\cite{ppg55,ppg28}.  For reference a line is plotted at 
$R_{d{\rm A}}$=1. The scaling systematic error is shown as a box on the 
left.}
\end{figure}

Figure~\ref{fig:rda} shows the nuclear modification factor 
$R^{\omega}_{d{\rm A}}$, defined as the ratio of the $\omega$ meson yields 
in $d$+Au interactions and $p+p$ interactions scaled by the number of 
binary collisions in $d$+Au, for minimum bias and central (0-20\%) $d$+Au 
events.  Precise definition of the $R_{d{\rm A}}$ and procedure to 
determine centrality in $d$+Au is given in sections IV.B and III.C 
in~\cite{ppg55}. 
We find $R^{\omega}_{d{\rm A}}$ to be 
$1.03 \pm 0.12^{\rm stat} \pm 0.21^{\rm sys}$ for minimum bias and 
$0.83 \pm 0.21^{\rm stat} \pm 0.17^{\rm sys}$ for central events, 
independent of $p_{\rm T}$.  The $R_{d{\rm A}}$ for two other neutral 
mesons ($\pi^{0}$ and $\eta$) measured by PHENIX~\cite{ppg55,ppg28} are also 
shown in Fig.~\ref{fig:rda}.  In all cases PHENIX observes that 
$R_{d{\rm A}}$ is close to one for $p_{\rm T} > 2$\,GeV/$c$ and flat out to 
the highest $p_{\rm T}$.  Similar behavior is seen in preliminary analysis 
of $K$ and $\phi$ mesons~\cite{victor,star-phi}.

%%%%%%%%%%%%%%%%%%%%%%%%%%%%%%%%%%%%%%%%%%%%%%%%%%%%%%%%%%%%%%  Fig. 3

\begin{figure}[tbh]
\includegraphics[width=1.0\linewidth]{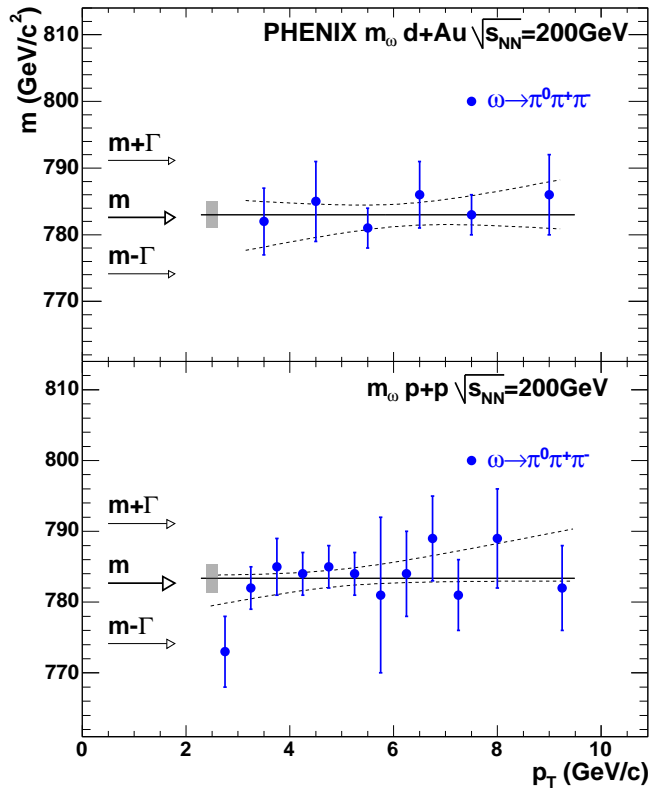}
\caption{Reconstructed $\omega$ mass measured in the
$\pi^{0}\pi^{+}\pi^{-}$ decay channel vs $p_{\rm T}$ in
(upper panel) $d$+Au and (lower panel) $p+p$ collisions at
$\sqrt{s_{NN}}=200$\,GeV.  Error bars show statistical
errors.  Straight lines show fits of the data to a
constant.  Dashed lines show values within $1\sigma$ of the best linear
fit to the data.  Systematic error on the fit value is shown with the box.  
PDG values for $\omega$ meson mass (m) and width ($\Gamma$) are shown with arrows on the
left~\cite{PDG}. \label{fig:mass}}
\end{figure}

Recent publications suggest that modifications to
the $\omega$ mass can be observed even in cold matter by studying not
only the electron decay channel~\cite{KEK,STAR_rho}, but also hadronic
channels~\cite{mass}.  For the hadronic decay modes presented here
PHENIX lacks acceptance at low $p_{\rm T}$ where the 
effect is expected to be the most prominent.  However, we do have
excellent mass resolution (20-25\,MeV) for the mixed neutral-charged particle decay
mode.  Figure~\ref{fig:mass} shows extracted values for the $\omega$ mass 
as a function of $p_{\rm T}$.
In the $p_{\rm T}$ range of the
measurement we observe no modification of the $\omega$ mass in either $d$+Au or $p+p$
collision systems at $\sqrt{s_{NN}}=200$\,GeV.

%%%%%%%%%%%%%%%%%%%%%%%%%%%%%%%%%%%%%%%%%%%%%%%%%%%%%%%%%%%%%%  Fig. 4

In summary we have presented the first measurement of $\omega$
production in $p+p$ and $d$+Au collisions at $\sqrt{s_{NN}} =
200$\,GeV.  The production cross section is measured in two
different decay modes with consistent results.  The $\omega/\pi^0$
ratio in $p+p$ collisions is found to be 
$0.85 \pm 0.05^{\rm stat} \pm 0.09^{\rm sys}$ and  
$0.94 \pm 0.08^{\rm stat} \pm 0.12^{\rm sys}$ 
in $d$+Au over the measured $p_{\rm T}$ range.  This agrees with 
previous measurements in hadronic collisions at lower $\sqrt{s}$.
The nuclear modification factor for
$\omega$ production in $d$+Au collisions is consistent with 1 and
$p_{\rm T}$ independent for $p_{\rm T} > 2$\,GeV/$c$ consistent with other meson
measurements.  No modifications to the
$\omega$ mass were observed in $p+p$ or $d$+Au collisions.  The
$\omega$ meson is also interesting as a probe of the hot nuclear
medium created in A+A collisions.  This measurement will serve as an
excellent baseline for measurements of $\omega$ production in A+A
in various decay channels.

%%%%%%%%%%%%%%%%%%%%%%%%%%%%%%%%%%%%%%%%%%%%%%%%%%%%%%% Acknowledgements
%\section{Acknowledgements}   % Run-3 short form for PRL

We thank the staff of the Collider-Accelerator and Physics
Departments at BNL for their vital contributions.  
We acknowledge support from 
the Department of Energy and NSF (U.S.A.), 
MEXT and JSPS (Japan), 
CNPq and FAPESP (Brazil), 
NSFC (China), 
IN2P3/CNRS, CEA, and ARMINES (France), 
BMBF, DAAD, and AvH (Germany), 
OTKA (Hungary), 
DAE and DST (India), 
ISF (Israel), 
KRF and KOSEF (Korea), 
RMIST, RAS, and RMAE (Russia), 
VR and KAW (Sweden), 
U.S. CRDF for the FSU, 
US-Hungarian NSF-OTKA-MTA, 
and US-Israel BSF.

%%%%%%%%%%%%%%%%%%%%%%%%%%%%%%%%%%%%%%%%%%%%%%%%%%%%%%%%%%%% References


\begin{references}

\bibitem{geist90} J.F.~Owens, Rev. Mod. Phys. {\bf 59}, 465 (1987)
\bibitem{loss} I.~Vitev, Phys. Lett. {\bf B562}, 36 (2003)
\bibitem{shadow} V.~Guzey, M.~Strikman, W.~Vogelsang Phys. Lett. {\bf B603}, 173 (2004)
\bibitem{aaa}  R.~Baier, D.~Schiff and B. Zakharov, Ann. Rev. Nucl. Part. Sci. {\bf 50}, 37 (2000)
\bibitem{recomb} R.~J.~Fries, J. Phys. {\bf G30}, S853 (2004)
\bibitem{vacuum} R.~Rapp and J.~Wambach, Adv. Nucl. Phys. {\bf 25}, 1 (2000)
\bibitem{phenix} K.~Adcox et al, (PHENIX Collaboration) Nucl.\ Instrum.\ Meth. {\bf A499}, 469 (2003)
\bibitem{ppg24} S.S.~Adler et al, (PHENIX Collaboration) Phys.\ Rev.\ Lett. {\bf 91}, 182301, (2003)
\bibitem{dau_totalxsec} S.~White, AIP Conf.\ Proc.\  {\bf 792}, 527 (2005)
\bibitem{ppg55} S.S.~Adler et al, (PHENIX Collaboration) nucl-ex/0611006
\bibitem{dalitz1} C.~Alff et al, Phys.\ Rev.\ Lett. {\bf 9} 325 (1962)
\bibitem{dalitz2} M.L.~Stevenson, Phys.\ Rev.\ {\bf 125}, 687 (1962)
\bibitem{ppg14} S.S.~Adler et al, (PHENIX Collaboration) Phys.\ Rev.\ Lett. {\bf 91}, 072301, (2003)
\bibitem{ppg51} S.S.~Adler et al, (PHENIX Collaboration) Phys.\ Rev.\ Lett. {\bf 96}, 202301, (2006)
\bibitem{victor} V.~Riabov (for the PHENIX Collaboration) Nucl.\ Phys. {\bf A774}, 735-738, (2006)
\bibitem{ppg28} S.S.~Adler et al, (PHENIX Collaboration) Phys.\ Rev.\ Lett. {\bf 91}, 072303, (2003)
\bibitem{pythia} PYTHIA 6.206, T.~Sjostrand et al, hep-ph/0108264 used with default parameters.
\bibitem{isr} M.~Diakonou et al, Phys Lett {\bf B89}, 432 (1980)
\bibitem{e706} L.~Apanasevich, FERMILAB-Pub-00/054-E (2000)
\bibitem{L3} M.~Acciarri et al., (L3 Collaboration) Phys.\ Lett.\ {\bf B393}, 465-476 (1997)
\bibitem{ALEPH} R.~Barate et al., (ALEPH Collaboration) Phys.\ Rep.\ {\bf 294}, 1-165 (1998)
\bibitem{OPAL} K.~Ackerstaff et al., (OPAL Collaboration) Eur.\ Phys.\ Jour.\ {\bf C5}, 411-437 (1998)
\bibitem{star-phi} X.~Cai (for the STAR Collaboration) Nucl.\ Phys.\ {\bf A774}, 485-488, (2006)
\bibitem{KEK} M.~Naruki et al, Phys.\ Rev.\ Lett. {\bf 96}, 092301, (2006)
\bibitem{STAR_rho} J.~Adams et al., (STAR Collaboration) Phys.\ Rev.\ Lett. {\bf 92}, 092301, (2004)
\bibitem{mass} D.~Trnka et al, Phys.\ Rev.\ Lett. {\bf 94}, 192303, (2005)
\bibitem{PDG} Particle Data Group. S.~Edelman, et.al. Phys.\ Lett. {\bf B592}, 1 (2004)

\end{references}
\end{document}